\input{aipcheck}

\documentclass[citeautoscript
    ,final            
  ,numberedheadings 
,sort&compress
  ]
  {aipproc}

\layoutstyle{8x11single}

\usepackage{amsmath,amssymb}
\usepackage{color}
\usepackage{natbib}

\begin{document}

\title{Highly frustrated spin-lattice models of magnetism and their quantum phase transitions: A microscopic treatment via the coupled cluster method}


\author{R.~F.~Bishop}{
  address={School of Physics and Astronomy, Schuster Building, The University of Manchester, Manchester, M13 9PL, UK}
}

\author{P.~H.~Y.~Li}{
  address={School of Physics and Astronomy, Schuster Building, The University of Manchester, Manchester, M13 9PL, UK}
}

\author{C.~E.~Campbell}{
  address={School of Physics and Astronomy, University of Minnesota, 116 Church Street SE, Minneapolis, Minnesota 55455, USA}
}

\begin{abstract}
  We outline how the coupled cluster method of microscopic quantum
  many-body theory can be utilized in practice to give highly accurate
  results for the ground-state properties of a wide variety of highly
  frustrated and strongly correlated spin-lattice models of interest
  in quantum magnetism, including their quantum phase transitions.
  The method itself is described, and it is shown how it may be
  implemented in practice to high orders in a systematically
  improvable hierarchy of (so-called LSUB$m$) approximations, by the use of
  computer-algebraic techniques.  The method works from the outset in
  the thermodynamic limit of an infinite lattice at all levels of
  approximation, and it is shown both how the ``raw'' LSUB$m$ results
  are themselves generally excellent in the sense that they converge
  rapidly, and how they may accurately be extrapolated to the exact
  limit, $m \rightarrow \infty$, of the truncation index $m$, which
  denotes the {\it only} approximation made.  All of this is
  illustrated via a specific application to a two-dimensional,
  frustrated, spin-half $J^{XXZ}_{1}$--$J^{XXZ}_{2}$ model on a
  honeycomb lattice with nearest-neighbor and next-nearest-neighbor
  interactions with exchange couplings $J_{1}>0$ and $J_{2} \equiv
  \kappa J_{1} > 0$, respectively, where both interactions are of the
  same anisotropic $XXZ$ type.  We show how the method can be used to
  determine the entire zero-temperature ground-state phase diagram of
  the model in the range $0 \leq \kappa \leq 1$ of the frustration
  parameter and $0 \leq \Delta \leq 1$ of the spin-space anisotropy
  parameter.  In particular, we identify a candidate quantum
  spin-liquid region in the phase space.
\end{abstract}

\maketitle


\section{INTRODUCTION}
The coupled cluster method (CCM) \cite{Bishop:1998_QMBT_coll} is one
of the most pervasive, most powerful, and most successful of all {\it
  ab initio} formalisms of quantum many-body theory.  It has probably
been applied to more systems in quantum field theory, quantum
chemistry, nuclear, subnuclear, condensed matter, and other areas of
physics than any other competing method.  The CCM has yielded
numerical results which are among the most accurate available for an
incredibly wide range of both finite and extended physical systems
defined on a spatial continuum.  These range from atoms and molecules
of interest in quantum chemistry, where the method has long been the
recognized ``gold standard'', to atomic nuclei; from the electron gas
to dense nuclear and baryonic matter; and from models in quantum
optics, quantum electronics, and solid-state optoelectronics to field
theories of strongly interacting nucleons and pions.  

This widespread success for both finite \cite{Bartlett:1989_ccm} and
extended \cite{Bishop:1991_TheorChimActa_QMBT} physical systems has
led to recent applications to corresponding quantum-mechanical systems
defined on an extended regular spatial lattice.  Such lattice systems
are nowadays the subject of intense theoretical study.  They include
many examples of systems characterized by novel ground states
which display {\it quantum order} in some region of the Hamiltonian
parameter space, delimited by critical values or {\it quantum critical
  points} (QCPs), which mark the corresponding {\it quantum phase
  transitions}.  The quantum critical phenomena often differ
profoundly from their classical counterparts, and the subtle
correlations present usually cannot easily be treated by standard
many-body techniques such as perturbation theory or mean-field
approximations.

A key challenge for modern quantum many-body theory has been to
develop microscopic techniques capable of handling both these novel
and more traditional systems.  Our recent work, in the field of
quantum magnetism, for example, shows that the CCM is clearly able to
bridge this divide.  We have shown how the systematic inclusion of
multispin correlations for a wide variety of quantum spin-lattice
problems can be efficiently implemented with the CCM
\cite{Fa:2004_QM-coll}.  The method is not restricted to bipartite
lattices or to non-frustrated systems, and can thus deal with problems
where many alternative techniques, such as the exact diagonalization
(ED) of small lattices or quantum Monte Carlo (QMC) simulations, are
faced with specific difficulties.

In this paper we illustrate the current power of the CCM to describe
accurately the properties of strongly interacting and highly
frustrated spin-lattice models of interest in quantum magnetism,
especially in two spatial dimensions.  The method itself is first
briefly reviewed in Sec.\ \ref{model_sec}, where we demonstrate how it
may readily be implemented to high orders in a specific,
systematically improvable, hierarchy ({\it viz}., a localized
lattice-animal-based subsystem, LSUB$m$, scheme) of approximations, by
the use of computer-algebraic techniques.  In order to demonstrate how
values for ground-state (GS) properties are obtained, using the CCM,
which are fully competitive with those from other state-of-the-art
methods, including the much more computationally intensive QMC
techniques in the relatively rare (unfrustrated) cases where the
latter can readily be applied, we apply it to a specific model of
current interest.  The model itself, which is a frustrated spin-half
($s=\frac{1}{2}$) antiferromagnet with nearest-neighbor (NN) $J_{1}>0$
and competing next-nearest-neighbor (NNN) $J_{2}>0$ exchange couplings
on the honeycomb lattice, both of the anisotropic $XXZ$ type, is
described in Sec.\ \ref{ccm_sec}.  Results for the model are presented
in Sec.\ \ref{results_sec}, where we demonstrate the ability of the
CCM to give an accurate description of the zero-temperature ($T=0$) GS
phase diagram of this model, which contains two independent control
parameter, {\it viz}., the frustration parameter $\kappa \equiv
J_{2}/J_{1}$, and the spin anisotropy parameter $\Delta$.  The raw
LSUB$m$ results themselves are shown to be generally excellent, and we
demonstrate explicitly both how they converge rapidly and can also be
accurately extrapolated in the truncation index to the exact limit, $m
\rightarrow \infty$.  We show in Sec.\ \ref{phase_sec} how the results
so obtained may be used to construct an accurate $T=0$ GS phase
diagram for this model.  Finally, in Sec.\ \ref{conclusion} we present
our conclusions.

\section{A HONEYCOMB LATTICE MODEL}
\label{model_sec}
Low-dimensional spin-lattice models of magnets exhibiting frustration,
due either to the underlying lattice geometry or to competing
interactions, have been the subject of intense study in recent years,
both at the theoretical level and via their experimental realizations
either in real materials or in ultracold atoms trapped in optical
lattices.  Their $T=0$ GS phase diagrams often differ profoundly from
their classical ($s \rightarrow \infty$) counterparts, exhibiting, for
example, such states without magnetic order as various valence-bond
crystalline (VBC) phases or quantum spin-liquid (QSL) states.

\begin{figure}
\mbox{
\includegraphics[height=3cm]{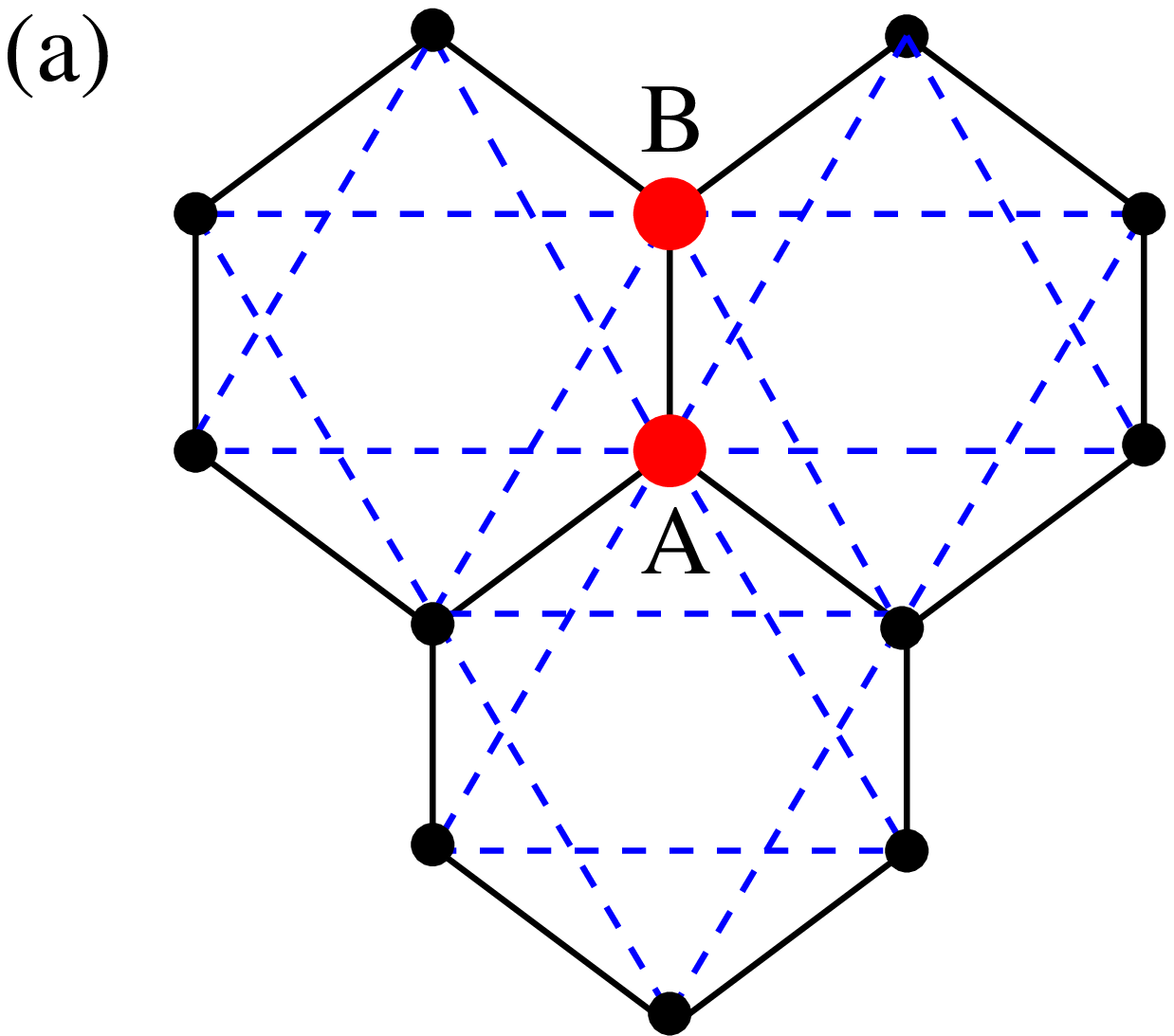}
  \quad
 \includegraphics[height=3cm]{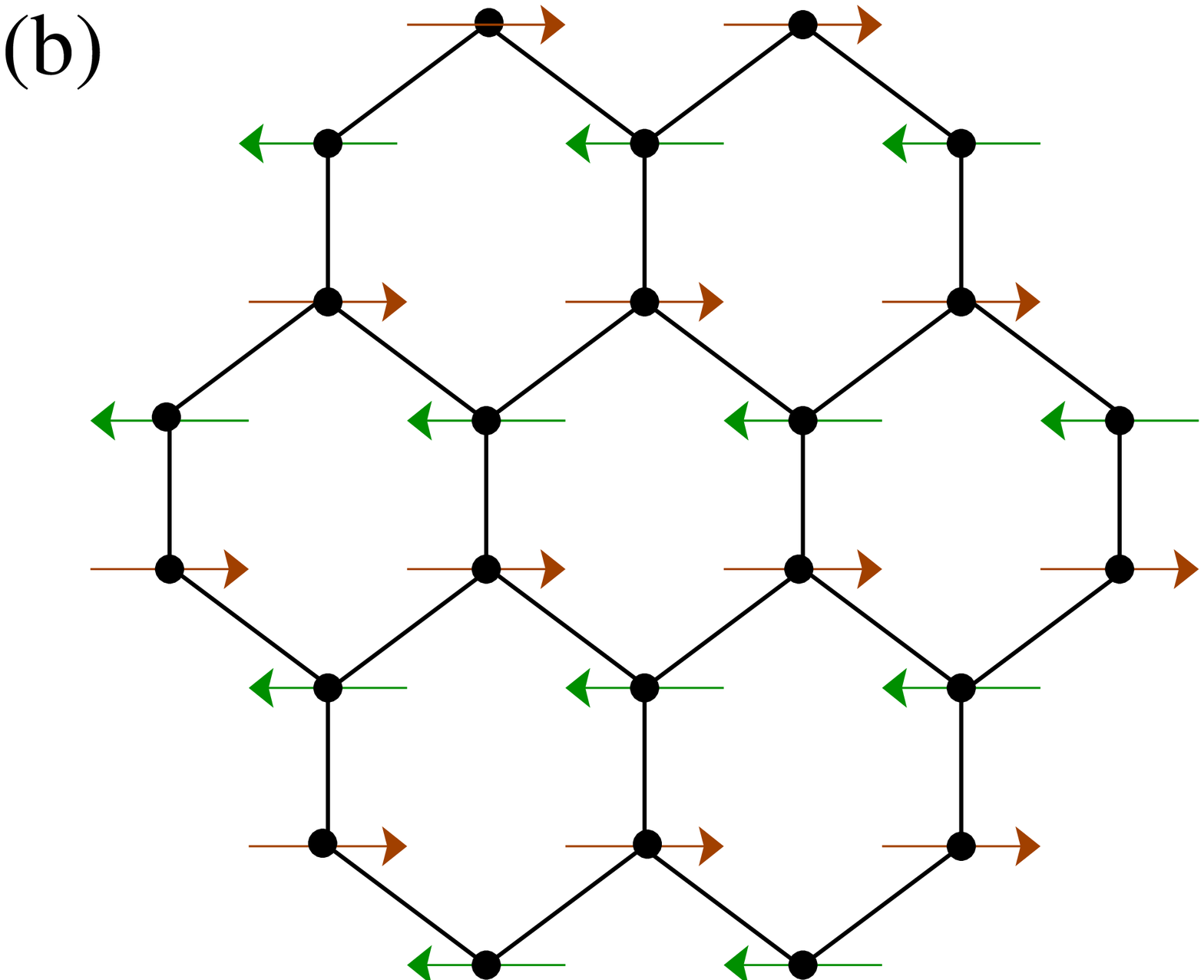}
  \quad
\includegraphics[height=3cm]{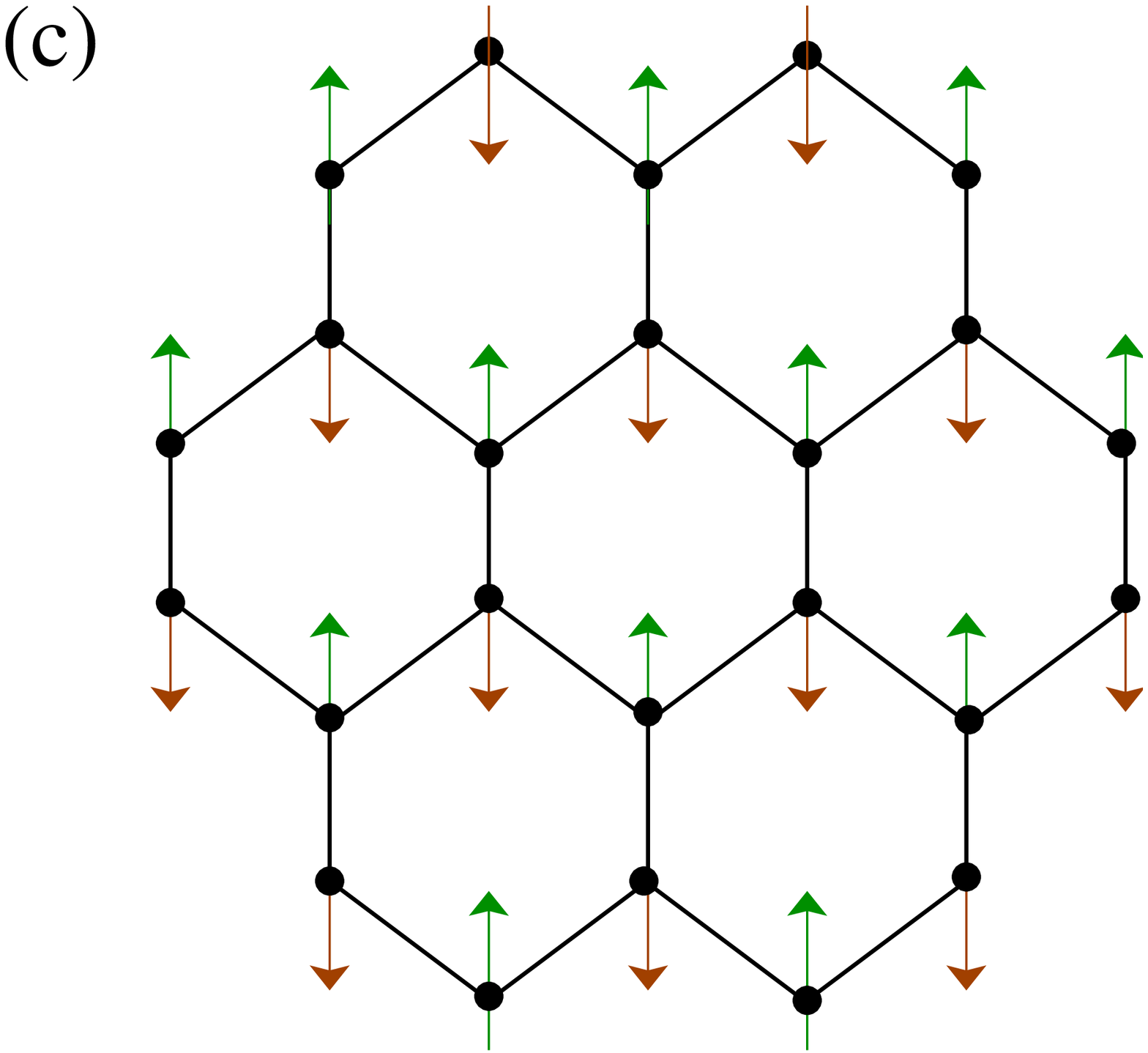}
  \quad 
\includegraphics[height=3cm]{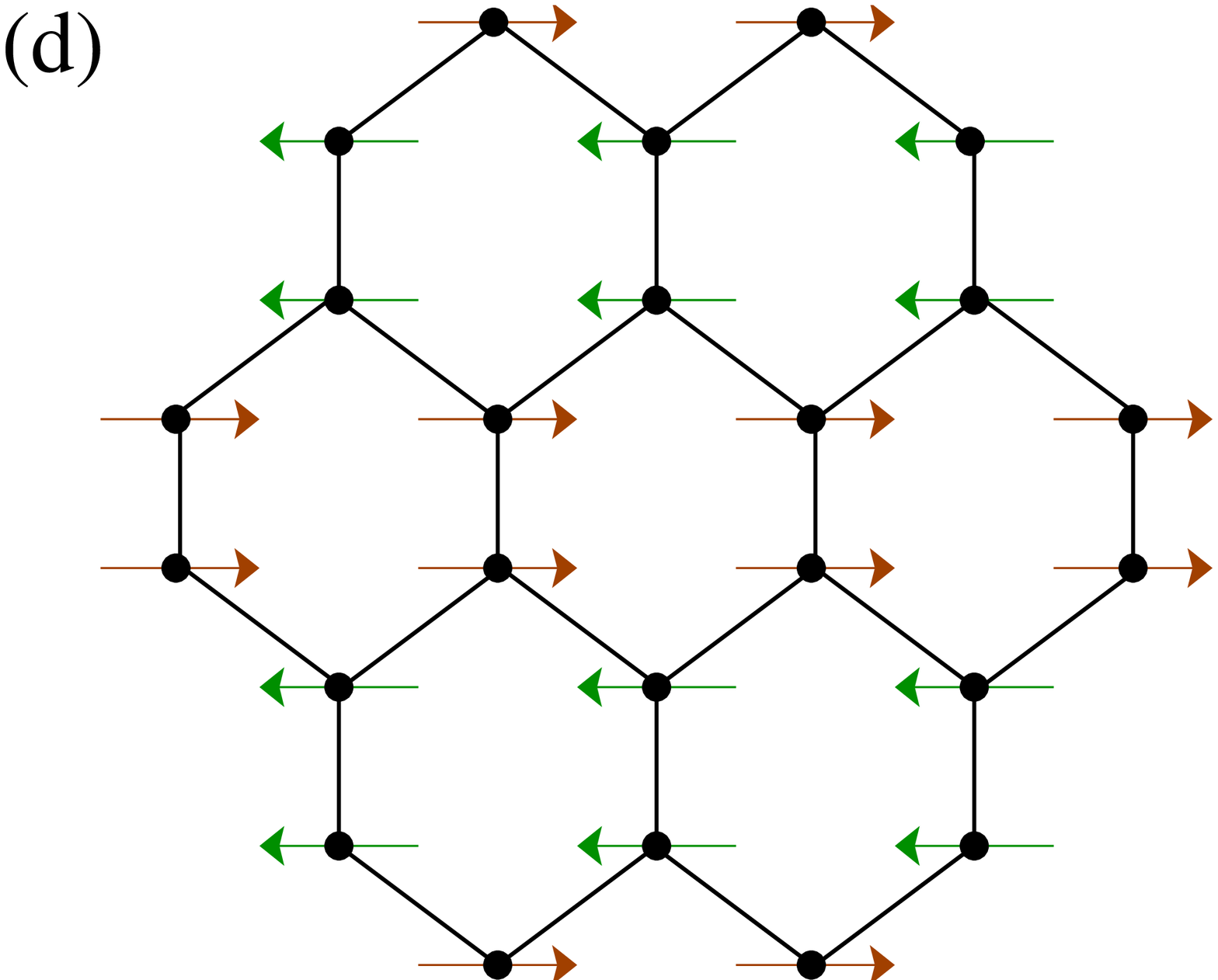}
}
  \caption{(Color online)  The $J^{XXZ}_{1}$--$J^{XXZ}_{2}$ model on the honeycomb lattice, showing (a) the bonds ($J_{1} \equiv$ ----- ; $J_{2} \equiv \textcolor{blue}{- - -}$) and the two sites (\textcolor{red}{$\bullet$}) A and B of the unit cell; (b) the N\'{e}el planar, N(p), state; (c) the N\'{e}el $z$-aligned, N($z$), state; and (d) the N\'{e}el-II planar, N-II(p), state.  The arrows represent the directions of the spins located on lattice sites \textcolor{black}{$\bullet$}.}
\label{model}
\end{figure}

Since quantum fluctuations of the order parameter destroy long-range
order and hence prevent most types of continuous symmetry breaking in
one-dimensional (1D) systems, even at $T=0$, 2D systems occupy a
special role for studying QPTs.  Since quantum fluctuations are
generally weaker for higher values of the spin quantum number $s$,
systems with $s=\frac{1}{2}$ typically exhibit the biggest differences
from classical behavior.  Furthermore, of all regular 2D lattices, one
with the lowest coordination number, $z=3$, is the honeycomb lattice.
Thus, good {\it prima facie} candidate systems for exhibiting novel
behavior are spin-half models on the honeycomb lattice, and as a
specific model that exhibits both frustration and anisotropy (in spin
space), we consider here the so-called $J^{XXZ}_{1}$--$J^{XXZ}_{2}$
model \cite{Li:2014_honey_XXZ}.  It is shown schematically in Fig.\
\ref{model}(a) and its Hamiltonian is given by
\begin{equation}
H = J_{1}\sum_{\langle i,j \rangle}(s^{x}_{i}s^{x}_{j}+s^{y}_{i}s^{y}_{j}+\Delta s^{z}_{i}s^{z}_{j}) + J_{2}\sum_{\langle\langle i,k \rangle\rangle}(s^{x}_{i}s^{x}_{k}+s^{y}_{i}s^{y}_{k}+\Delta s^{z}_{i}s^{z}_{k})\,, \label{H_honey_XXZ}
\end{equation}
where $\langle i,j \rangle$ and $\langle\langle i,k \rangle\rangle$
denote NN and NNN pairs of spins respectively, and the respective
sums count each bond once and once only; and ${\bf
  s}_{i}=(s^{x}_{i},s^{y}_{i},s^{z}_{i}$) is the $s=\frac{1}{2}$ spin 
operator on the $i$th site of the honeycomb lattice.  We shall be
interested in the thermodynamic limit of an infinite lattice ($N
\rightarrow \infty$, where $N$ is the number of lattice sites).

The model of Eq.\ (\ref{H_honey_XXZ}) interpolates continuously
between the two cases where both NN and NNN exchange couplings have
either an isotropic Heisenberg ($XXX$) form when $\Delta=1$ or an
isotropic $XY$ ($XX$) form when $\Delta=0$.  We shall be interested in
the case where both bonds are antiferromagnetic in nature (i.e., when
$J_{1}>0$ and $J_{2}>0$), so that they act to frustrate one another.
With no further loss of generality we henceforth put $J_{1} \equiv 1$
to set the overall energy scale, and we study the model in the range
$0 \leq \kappa \leq 1$ of the frustration parameter $\kappa \equiv
J_{2}/J_{1}$, and $0 \leq \Delta \leq 1$ of the spin anisotropy
parameter.  Although both limiting isotropic $s=\frac{1}{2}$ models on
the honeycomb lattice have been well studied in the past (see, Refs.\
\cite{Varney:2011_honey_XY,Varney:2012_honey_XY,Zhu:2013_honey_XY,Carrasquilla:2013_honey_XY,Ciolo:2014_honey_XY,Oitmaa:2014_honey_XY,Bishop:2014_honey_XY}
for the $\Delta=0$ $XX$ model and Refs.\
\cite{Rastelli:1979_honey,Fouet:2001_honey,Mulder:2010_honey,
  Ganesh:2011_honey,Clark:2011_honey,Albuquerque:2011_honey,
  Mosadeq:2011_honey,Oitmaa:2011_honey,Mezzacapo:2012_honey,Li:2012_honey_full,Bishop:2012_honeyJ1-J2,RFB:2013_hcomb_SDVBC,Ganesh:2013_honey_J1J2mod-XXX,Zhu:2013_honey_J1J2mod-XXZ,Gong:2013_J1J2mod-XXX,Yu:2014_honey_J1J2mod-XXZ}
for the $\Delta=1$ $XXX$ model, there is still no overall consensus
for either model for its respective complete $T=0$ GS phase diagram in
the range of values of $\kappa$ and $\Delta$ under study.  What is
agreed, however, is that although the two limiting models share
exactly the same $T=0$ GS phase diagram in the classical ($s
\rightarrow \infty$) case \cite{Rastelli:1979_honey,Fouet:2001_honey},
their $s=\frac{1}{2}$ counterparts differ significantly.  For this
reason alone, a complete study of the $T=0$ GS phase diagram of the
$s=\frac{1}{2}$ model of Eq.\ (\ref{H_honey_XXZ}) on the honeycomb
lattice is of clear interest.

There is broad agreement from various theoretical studies that whereas
both classical ($s \rightarrow \infty$) $XX$ and $XXX$ models have
N\'{e}el ordering for $\kappa < \kappa_{{\rm cl}} =\frac{1}{6}$, their
$s=\frac{1}{2}$ counterparts both retain N\'{e}el order out to larger
values $\kappa_{c_{1}} \approx 0.2$.  This finding is completely
consistent with the general observation that quantum fluctuations tend
to favor collinear forms of magnetic order over noncollinear ones
since, in the classical cases, for $\kappa > \kappa_{{\rm cl}}$ the GS
phase comprises an infinitely degenerate family of states with spiral
magnetic order (and see Refs.\
\cite{Rastelli:1979_honey,Fouet:2001_honey}).  These spirally-ordered
noncollinear states are very fragile against quantum fluctuations, and
there is by now a broad consensus in the literature that neither the 
$s=\frac{1}{2}$ $XX$ or $XXX$ model has a stable $T=0$ GS phase with
noncollinear spiral ordering for any value of $\kappa$ in the range $0
\leq \kappa \leq 1$ under study.  On the other hand, as $\kappa
\rightarrow \infty$, both models reduce to Heisenberg antiferromagnets
(HAFs) on two independent triangular lattices, for each of which one
knows that the stable GS phase is one where the spins are arranged on
three sublattices with relative 120$^{\circ}$ ordering.  Whether such
a state is stable against the imposition of NN $J_{1}$ exchange
coupling for large but finite values of $\kappa$, or whether it then
transforms continuously to a spiral state with a given pitch angle for
a specific finite value of $\kappa$, is still unknown.  What is
broadly agreed, on the other hand, is that any such state only exists
for values $\kappa > 1$.

The most interesting region for both the $s=\frac{1}{2}$ $XX$ and
$XXX$ models is when $\kappa \gtrsim 0.2$.  Thus, we know that novel
quantum phases often emerge from classical models which have an
infinitely degenerate family of GS phases in some region of phase
space, as is the case here for the classical $XX$ and $XXX$ models for
$\kappa > \kappa_{{\rm cl}}=\frac{1}{6}$.  What is typically then
found is that quantum fluctuations lift this (accidental) GS
degeneracy, either wholly or partially, by the well-known {\it order
  by disorder} mechanism \cite{Villain:1977_ordByDisord,
  Villain:1980_ordByDisord}.  Either one or several members,
respectively, of the classical family are then favored as the quantum
GS phase.  For the present $XXX$ model on the honeycomb lattice, for
example, it has been shown \cite{Mulder:2010_honey} that to leading
order, $O(1/s)$, spin-wave fluctuations lift the degeneracy in favor
of specific wave vectors, leading to spiral order by disorder.

On the other hand, we know too that quantum fluctuations generally
favor collinear ordering over noncollinear ordering, as mentioned
above.  Hence, one may easily intuit that the strong quantum
fluctuations present in the $s=\frac{1}{2}$ models might melt the
spiral order for a wide range of values of $\kappa$ in favor of some
collinear state.  One such clear collinear candidate state is actually
among the infinitely degenerate family of ground states at the
classical critical point $\kappa=\frac{1}{2}$, at which the closed
contours of values of the spiral wave vector, all of which minimize
the classical GS energy for a given value of $\kappa$, change
character \cite{Mulder:2010_honey}.  This special collinear state
among the infinite family of $\kappa=\frac{1}{2}$ ground states is the
so-called N\'{e}el-II state.  It is characterized by having all NN
bonds along any one of the three equivalent honeycomb lattice
directions as being ferromagnetic (i.e., with spins parallel), while
those along the remaining two directions are antiferromagnetic (i.e.,
with spins antiparallel), as illustrated in Fig.\ \ref{model}(d), for example.

In the extreme $s=\frac{1}{2}$ quantum limit one may also expect
quantum fluctuations to destroy completely the magnetic order in any
(collinear or noncollinear) quasiclassical state in some region or
other of the $T=0$ GS phase space.  Just such paramagnetic states have
been found by using various theoretical techniques, for both the
$s=\frac{1}{2}$ $XX$ and $XXX$ models on the honeycomb lattice, in the
interesting regime $0.2 \lesssim \kappa \lesssim 0.4$ where, however,
the least consensus exists for either model.  For the $s=\frac{1}{2}$
$XX$ model, for example, the N\'{e}el $xy$ planar [N(p)] ordering that
exists for $\kappa < \kappa_{c_{1}} \approx 0.2$ is predicted by
different techniques to give way either to a GS phase with N\'{e}el
$z$-aligned [N($z$)] order
\cite{Zhu:2013_honey_XY,Bishop:2014_honey_XY} or to one with a QSL
nature \cite{Varney:2011_honey_XY,Carrasquilla:2013_honey_XY} in a
range $\kappa_{c_{1}} < \kappa < \kappa_{c_{2}} \approx 0.4$.  By
contrast, for the $s=\frac{1}{2}$ $XXX$ model, the N\'{e}el order that
exists for $\kappa < \kappa_{c_{1}}$ is variously predicted to give
way either to a GS phase with plaquette valence-bond crystalline (PVBC)
order \cite{Albuquerque:2011_honey,
  Mosadeq:2011_honey,Li:2012_honey_full,Bishop:2012_honeyJ1-J2,RFB:2013_hcomb_SDVBC,Ganesh:2013_honey_J1J2mod-XXX,Zhu:2013_honey_J1J2mod-XXZ}
or to a QSL state
\cite{Clark:2011_honey,Mezzacapo:2012_honey,Gong:2013_J1J2mod-XXX,Yu:2014_honey_J1J2mod-XXZ}
in the corresponding range $\kappa_{c_{1}} < \kappa < \kappa_{c_{2}}$.

In the range ($1 >$) $\kappa > \kappa_{c_{2}}$ there is broad agreement
that for both models there is a strong competition to form the GS
phase between states with collinear N\'{e}el-II $xy$ planar [N-II(p)]
order and staggered-dimer valence-bond crystalline (SDVBC) order,
which lie very close in energy to one another.  Both of these states
break the lattice rotational symmetry in the same way, and are
correspondingly threefold-degenerate.  Some theoretical treatments
also favor a further QCP at $\kappa_{c_{3}} > \kappa_{c_{2}}$, at
which a transition occurs between a GS phase with SDVBC ordering for
$\kappa_{c_{2}} < \kappa < \kappa_{c_{3}}$, possibility mixed in some
or all of this regime with N-II(p) ordering, to one with N-II(p)
ordering alone for $\kappa > \kappa_{c_{3}}$.  It is interesting to
note in this context that alternative techniques such as the ED and
density-matrix renormalization group (DMRG) methods, both of which are
restricted to lattices with a finite number $N$ of lattice sites, find
it particularly difficult to distinguish between the N-II(p) and SDVBC phases
in the regime $\kappa > \kappa_{c_{2}}$ in the thermodynamic limit $N
\rightarrow \infty$ in which we are interested, for which finite-size
scaling is required, especially for the $XX$ model.  It is thus
particularly valuable to use a size-extensive method such as the CCM
used here, which works from the outset in the $N \rightarrow \infty$
limit at every level of LSUB$m$ approximation.  Since such LSUB$m$
approximations form well-defined hierarchies, as explained in Sec.\
\ref{ccm_sec}, the only final extrapolation needed by us is to the
exact ($m \rightarrow \infty$) limit in the truncation index $m$.
Furthermore, at the highest level of approximation feasible with
available computational resources, results for physical quantities are
often already very well converged, as our specific results in Sec.\
\ref{results_sec} for the $s=\frac{1}{2}$ $J^{XXZ}_{1}$--$J^{XXZ}_{2}$
model of Eq.\ (\ref{H_honey_XXZ}) on the honeycomb lattice will show.

\section{THE COUPLED CLUSTER METHOD}
\label{ccm_sec}
Since the CCM is well documented in the literature (see, e.g., Refs.\
\cite{Bishop:2012_honeyJ1-J2,RFB:2013_hcomb_SDVBC,Bishop:2014_honey_XY,Bishop:1987_ccm,Arponen:1991_ccm,Bishop:1991_TheorChimActa_QMBT,Bishop:1998_QMBT_coll,Bishop:1991_XXZ_PRB44,Zeng:1998_SqLatt_TrianLatt,Farnell:2002_1D,Fa:2004_QM-coll,Bi:2008_PRB_J1xxzJ2xxz})
we present only a brief overview of its key features here.  Any CCM
calculation starts with the choice of a suitable model state (or
reference state), $|\Phi \rangle$, on top of which the quantum
correlations present in the exact GS phase under study can be
systematically incorporated later, as we describe below.  For the
present model we use each of the N(p), N($z$), and N-II(p) states
shown schematically in Figs.\
\ref{model}(b)--\ref{model}(d).

Once a model state $|\Phi \rangle$ is chosen, the exact GS ket- and
bra-state wave functions that satisfy the corresponding
Schr\"{o}dinger equations,
\begin{equation}
H|\Psi\rangle = E|\Psi\rangle\,; \quad \langle \tilde{\Psi}|H = E\langle \tilde{\Psi}|\,, \quad  \label{ket_bra_eqs}
\end{equation}
are parametrized as 
\begin{equation}
|\Psi \rangle = {\rm e}^{S} |\Phi \rangle\,; \quad \langle \tilde{\Psi}| = \langle\Phi|\tilde{S}{\rm e}^{-S}\,, \label{para_ket_bra_eqs}
\end{equation}
where we use the intermediate normalization scheme for $|\Psi\rangle$,
such that $\langle\Phi|\Psi\rangle = \langle\Phi|\Phi\rangle \equiv 1$, and
then for $\langle \tilde{\Psi}|$ choose its normalization such that
$\langle\tilde{\Psi}|\Psi\rangle = 1$.  The correlation operators
$S$ and $\tilde{S}$ are decomposed in terms of exact sets of
multiparticle, multiconfigurational creation and destruction operators,
$C^{+}_{I}$ and $C^{-}_{I} \equiv
(C^{+}_{I})^{\dagger}$, respectively, as
\begin{equation}
S=\sum_{I \neq 0}{\cal S}_{I}C^{+}_{I}\,; \quad \tilde{S} = 1 + \sum_{I \neq 0}\tilde{{\cal S}}_{I}C^{-}_{I}\,,   \label{corr_operators}
\end{equation}
where $C^{+}_{0} \equiv 1$, the identity operator, and $I$ is a set
index describing a complete set of single-particle configurations for
all of the particles.  The reference state $|\Phi \rangle$ thus acts as
a fiducial (or cyclic) vector, or generalized vacuum state, with respect to
the complete set of creation operators $\{C^{+}_{I}\}$, which are hence required to
satisfy the conditions $\langle \Phi|C^{+}_{I} = 0 =
C^{-}_{I}|\Phi\rangle, \forall I \neq 0$.

In order to consider each site on the spin lattice to be equivalent to
all others, whatever the choice of state $|\Phi \rangle$, it is
convenient to form a passive rotation of each spin so that in its own
local spin-coordinate frame it points in the downward, (i.e., negative
$z$) direction.  Clearly, such choices of local spin-coordinate frames
leave the basic SU(2) spin commutation relations unchanged, but have
the beneficial effect that the $C^{+}_{I}$ operators can be expressed as
products of single-spin raising operators $s^{+}_{k} \equiv s^{x}_{k}
+ is^{y}_{k}$, such that $C^{+}_{I} \equiv s^{+}_{k_{1}}s^{+}_{k_{2}}\cdots
s^{+}_{k_{n}};\;n=1,2,\cdots,2sN$.

The complete set of multiparticle correlation coefficients
$\{{\cal S}_{I},{\tilde{\cal S}}_{I}\}$ may now be evaluated by extremizing the
energy expectation value
$\bar{H}\equiv\langle\tilde{\Psi}|H|\Psi\rangle=\langle\Phi|{\tilde{S}}{\rm e}^{-S}H{\rm e}^{S}|\Phi\rangle$, with respect to each of
them, $\forall I \neq 0$.  Variation with respect to each coefficient
${\tilde{\cal S}}_{I}$ yields the coupled set of nonlinear equations,
\begin{equation}
\langle\Phi|C^{-}_{I}{\rm e}^{-S}H{\rm e}^{S}|\Phi\rangle=0\,, \quad \forall I \neq 0\,,  \label{nonlinear_eq}
\end{equation}
for the coefficients $\{{\cal S}_{I}\}$, while variation with respect to
each coefficient ${\cal S}_{I}$ yields the corresponding set of linear
equations,
\begin{equation}
\langle\Phi|\tilde{S}({\rm e}^{-S}H{\rm e}^{S} -
E)C^{+}_{I}|\Phi\rangle=0\,, \quad \forall I \neq 0\,,   \label{ket_linearEqs}
\end{equation}
for the coefficients $\{{\tilde{\cal S}}_{I}\}$, once the coefficients
$\{{\cal S}_{I}\}$ have been calculated from Eq.~(\ref{nonlinear_eq}),
and where in Eq.~(\ref{ket_linearEqs}) we have used
Eqs.~(\ref{ket_bra_eqs}) and (\ref{para_ket_bra_eqs}) to introduce
the GS energy $E$.

Up till now everything has been exact.  In practice, of course,
approximations need to be introduced, and these are made within the
CCM by restricting the set of indices $\{I\}$ retained in the
expansions of Eq.~(\ref{corr_operators}) for the otherwise exact
correlation operators $S$ and $\tilde{S}$.  One such specific
hierarchical scheme, {\it viz}., the LSUB$m$ scheme, is described
below.  It is important to realize, however, that no further
approximations are made.  In particular, the method is guaranteed by
the use of the exponential parametrizations in
Eq.~(\ref{para_ket_bra_eqs}) to be size-extensive at every level of
truncation, and hence we work from the outset in the $N \rightarrow
\infty$ limit.  Similarly, the important Hellmann-Feynman theorem is
also exactly obeyed at every level of truncation.  Lastly, when the
similarity-transformed Hamiltonian ${\rm e}^{-S}H{\rm e}^{S}$ in
Eqs.~(\ref{nonlinear_eq}) and (\ref{ket_linearEqs}) is expanded in
powers of $S$ using the well-known nested commutator expansion, the
fact that $S$ contains only spin-raising operators not only guarantees
that all terms are linked, but also that the otherwise infinite
expansion actually terminates at a finite order, so that no further
approximations are needed.

Once an approximation has been chosen and the retained coefficients
$\{{\cal S}_{I},{\tilde{\cal S}}_{I}\}$ have been calculated from
Eqs.~(\ref{nonlinear_eq}) and (\ref{ket_linearEqs}), any GS quantity
can, in principle, be calculated.  For example, the GS energy $E$ can
be calculated in terms of the coefficients $\{{\cal S}_{I}\}$ alone, as
$E=\langle\Phi|{\rm e}^{-S}H{\rm e}^{S}|\Phi\rangle$, while the average
on-site GS magnetization (or magnetic order parameter) $M$ needs both
sets $\{{\cal S}_{I}\}$ and $\{{\tilde{\cal S}}_{I}\}$ for its evaluation as
$M = -\frac{1}{N}\langle\Phi|\tilde{S}{\rm
  e}^{-S}\sum^{N}_{k=1}s^{z}_{k}{\rm e}^{S}|\Phi\rangle$, in terms of
the rotated local spin-coordinate frames defined above.

Thus, the {\it only} approximation made in the CCM is to truncate the
set of indices $\{I\}$ in the expansions of the correlation operators
$S$ and $\tilde{S}$.  We use here the well-studied LSUB$m$ scheme
\cite{Bishop:2012_honeyJ1-J2,RFB:2013_hcomb_SDVBC,Bishop:2014_honey_XY,Bishop:1991_XXZ_PRB44,Zeng:1998_SqLatt_TrianLatt,Farnell:2002_1D,Fa:2004_QM-coll,Bi:2008_PRB_J1xxzJ2xxz}
in which, at the $m$th level of approximation, one retains all
multispin-flip configurations $\{I\}$ defined over no more than $m$
contiguous lattice sites.  Such cluster configurations are defined to
be contiguous if every site is NN to at least one other.  The number,
$N_{f}$, of such fundamental configurations is reduced by exploiting
the space- and point-group symmetries and any conservation laws that
pertain to the Hamiltonian and the model state being used.  Even so,
$N_{f}$ increases rapidly with increasing LSUB$m$ truncation index
$m$, and it becomes necessary to use massive parallelization together
with supercomputing resources
\cite{Zeng:1998_SqLatt_TrianLatt},\footnote{We use the program package
  CCCM of D.~J.~J. Farnell and J.~Schulenburg, see
  http://www-e.uni-magdeburg.de/jschulen/ccm/index.html.} to derive
and solve the corresponding coupled sets of CCM equations
(\ref{nonlinear_eq}) and (\ref{ket_linearEqs}). For example, we
have finally $N_{f}=818\,300$ for the N-II(p) reference state at the
LSUB12 level.

Finally, as a last step, we need to extrapolate the approximate LSUB$m$ results to the
limit $m \rightarrow \infty$ where the CCM becomes exact.  For the GS
energy per spin, $e \equiv E/N$, we use the well-tested
extrapolation scheme
\cite{Farnell:2002_1D,Bishop:2012_honeyJ1-J2,RFB:2013_hcomb_SDVBC,Bishop:2014_honey_XY,Bi:2008_PRB_J1xxzJ2xxz,Fa:2004_QM-coll},
\begin{equation}
e(m) = e_{0}+e_{1}m^{-2}+e_{2}m^{-4}\,,   \label{E_extrapo}
\end{equation}
where results with
$m=\{6,8,10,12\}$ are employed for the N(p) and N-II(p) states used as
model state, and with $m=\{4,6,8,10\}$ for the N($z$) state.  For the
magnetic order parameter of systems near a QCP an appropriate
extrapolation rule is the ``leading power-law'' scheme
\cite{RFB:2013_hcomb_SDVBC,Bishop:2014_honey_XY},
\begin{equation}
M(m)=c_{0}+c_{1}(1/m)^{c_{2}}\,,  \label{M_extrapo_nu}
\end{equation}
which we use here for the LSUB$m$
results based on the N($z$) state with $m=\{4,6,8,10\}$.  An
alternative well-tested scheme for systems with strong frustration or
where the order in question is zero or close to zero
\cite{Bishop:2012_honeyJ1-J2,RFB:2013_hcomb_SDVBC,Bishop:2014_honey_XY}
is 
\begin{equation}
M(m)=d_{0}+d_{1}m^{-1/2}+d_{2}m^{-3/2}\,,  \label{M_extrapo_frustrated}
\end{equation}
when the leading exponent
$c_{2}$ in Eq.\ (\ref{M_extrapo_nu}) has been empirically found to be close to 0.5, as is the
case here for results based on both the N(p) and N-II(p) model states with $m=\{6,8,10,12\}$.

\section{RESULTS}
\label{results_sec}
We now firstly present our CCM extrapolated (LSUB$\infty$) results for
the GS energy per spin, $E/N$, and magnetic order parameter, $M$,
using the extrapolation schemes described above in Sec.\
\ref{ccm_sec}.  For both quantities we present three different curves
for each value of the anisotropy parameter $\Delta$ shown,
corresponding respectively to calculations based on the N(p), N($z$),
and N-II(p) states as our chosen CCM model state.

\begin{figure}[!t]
  \includegraphics[angle=270,width=9cm]{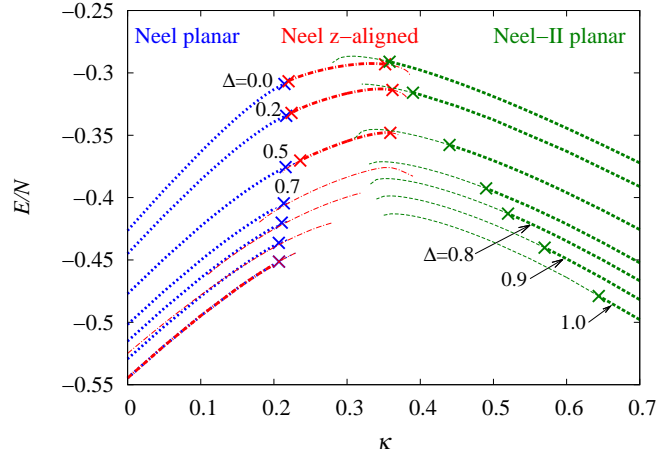}
  \caption{(Color online) The GS energy per spin $E/N$ versus the
    frustration parameter $\kappa \equiv J_{2}/J_{1}$ for the
    spin-$\frac{1}{2}$ $J^{XXZ}_{1}$--$J^{XXZ}_{2}$ model on the
    honeycomb lattice (with $J_{1}=1$), for various values of the
    anisotropy parameter $\Delta = 0.0, 0.2, 0.5, 0.7, 0.8, 0.9, 1.0$
    (from top to bottom, respectively).  We show extrapolated CCM
    LSUB$\infty$ results (see text for details) based on the N\'{e}el
    planar, N\'{e}el $z$-aligned, and N\'{e}el-II planar model states, respectively.  The times ($\times$) symbols mark the
    points where the respective extrapolations for the order parameter
    have $M \rightarrow 0$, and the unphysical portions of the
    solutions are shown by thinner lines (see text for details).}
\label{E}
\end{figure}

Results for the GS energy obtained in this way are shown in Fig.\ \ref{E}.  
A particularly noteworthy feature of the curves shown is that they all
exhibit {\it termination points}.  Thus, the N(p) curves all end at
corresponding upper termination points, while the N-II(p) curves end
at corresponding lower termination points.  The intermediate N($z$)
curves end at both corresponding lower and upper termination points.
In each case the respective termination points relate to those
points beyond which real solutions for the CCM multiconfigurational
correlation coefficients $\{{\cal S}_{I}\}$ cease to exist in the
LSUB$m$ approximation with the highest value of the truncation index
$m$ used, for the particular extrapolated curve shown.  Such
termination points of LSUB$m$ solutions are both well understood and well
documented in the literature (see, e.g., Refs.\
\cite{Fa:2004_QM-coll,Bishop:2014_honey_XY,Bishop:2012_honeyJ1-J2,RFB:2013_hcomb_SDVBC}).
They are simply approximate manifestations of a corresponding QCP in
the system, beyond which the order associated with the model state
being employed melts.  As would then be expected, we find for a given
value of $\Delta$ that as the index $m$ is increased the range of
values of $\kappa$ for which the LSUB$m$ equations have real solutions
becomes narrower.  Eventually, as $m \rightarrow \infty$, each
termination point then becomes the respective exact QCP.  Clearly,
from what has just been explained, real LSUB$m$ solutions with a fixed
finite value of $m$ can hence also exist in regions where the
corresponding magnetic order is destroyed (i.e., where $M<0$).

\begin{figure}[!b]
  \includegraphics[angle=270,width=9cm]{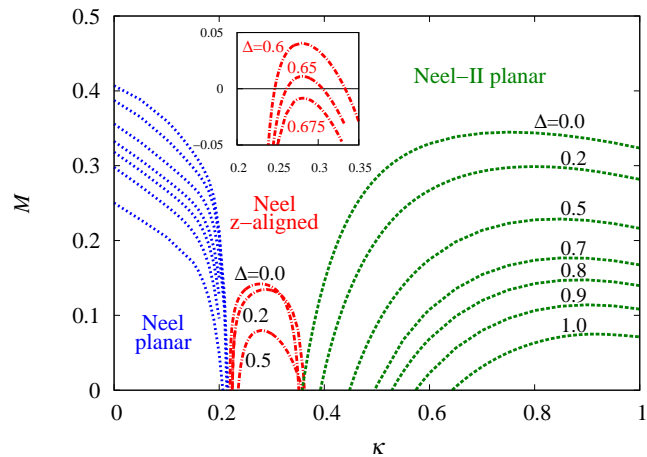}
  \caption{(Color online) The GS magnetic order parameter $M$ versus the frustration parameter $\kappa \equiv J_{2}/J_{1}$ for the spin-$\frac{1}{2}$ $J^{XXZ}_{1}$--$J^{XXZ}_{2}$ model on the honeycomb lattice (with $J_{1}>0$) for various values of the anisotropy parameter $\Delta=0.0,0.2,0.5,0.7,0.8,0.9,1.0$ (from top to bottom, respectively).  We show extrapolated CCM LSUB$\infty$ results (see text for details) based on the N\'{e}el planar, N\'{e}el $z$-aligned, and N\'{e}el-II planar states as CCM model states, respectively.}
\label{M}
\end{figure}

Corresponding sets of curves to those shown in Fig.\ \ref{E} for the
GS energy per spin, $E/N$, are shown in Fig.\ \ref{M} for the magnetic
order parameter, $M$.
In Fig.\ \ref{E} we show by times ($\times$) symbols those points on
each curve where $M=0$, as determined from the corresponding
extrapolated LSUB$\infty$ curve in Fig.\ \ref{M}.  In Fig.\ \ref{E} we
also denote by thinner lines those portions of the curves which are
``unphysical'' in the sense that $M<0$, by contrast with the
corresponding ``physical'' regions where $M>0$, which pertain to those
portions of the curves denoted by thicker lines.

We can immediately draw several conclusions from the results shown in
Figs.\ \ref{E} and \ref{M}.  Firstly, it is clear that N(p) order is
present, for all values of $\Delta$ shown, below a lower critical
value, $0 < \kappa < \kappa_{c_{l}}(\Delta)$.  Furthermore,
$\kappa_{c_{l}}$ depends only very weakly on $\Delta$, taking the
value $\kappa_{c_{l}}(\Delta) \approx 0.21$.  Secondly, we observe both 
that N($z$) order is present within a rather narrow range of values
around $\kappa \approx 0.3$ for $\Delta \lesssim 0.66$, but that it
becomes unstable for $\Delta \gtrsim 0.66$.  Thirdly, it is also clear
that N-II(p) order is present, for all values of $\Delta$ shown, above
some upper critical value, $\kappa_{c_{u}}(\Delta) < \kappa$ $(< 1)$,
where $\kappa_{c_{u}}(\Delta)$ increases monotonically with $\Delta$.
Fourthly, it is particularly clear from Fig.\ \ref{M} that the GS
phases with N(p) and N($z$) order melt at (or very close to) the {\it
  same} value $\kappa_{c_{l}}(0)$ for $\Delta=0$, but as $\Delta$ is
increased a very narrow region (in $\kappa$) opens up between these
two phases in which the GS phase has neither of these orderings.
Finally, Fig.\ \ref{M} similarly shows that although the two GS phases
with N($z$) and N-II(p) order also melt at (or very close to) the {\it
  same} value $\kappa_{c_{u}}(0)$ for $\Delta=0$, as $\Delta$ is
increased a GS phase with neither of these forms of order opens up
between them.  The range (in $\kappa$) of stability of this
intermediate phase increases monotonically with $\Delta$.

We now turn to the issue of what might be the nature of the
remaining GS phases outside the regimes of stability of the
quasiclassical N(p), N($z$), and N-II(p) phases, as discussed above.
Once we have identified any possible candidate phase with a specific
form of ordering, described by a suitable operator $\hat{O}$, a very
convenient way to test for the relative stability of a GS phase built on a
given CCM model state against that new form of ordering is to consider its
linear response to an imposed perturbation with a corresponding field
operator, $F = \delta\,\hat{O}$, added to the original system
Hamiltonian [i.e., of Eq.\ (\ref{H_honey_XXZ}) for the present case],
where $\delta$ is a (positive) infinitesimal.  The perturbed energy
per spin, $e(\delta) \equiv E(\delta)/N$, is then calculated at
various LSUB$m$ levels of approximation based on the CCM model state
whose stability is being investigated, for the infinitesimally
perturbed Hamiltonian $H + F$.  The corresponding susceptibility of
the system to this perturbation is then defined, as usual, (and see,
e.g., Refs.\
\cite{Bishop:2014_honey_XY,Bishop:2012_honeyJ1-J2,RFB:2013_hcomb_SDVBC})
as
\begin{equation}
\chi
\equiv - \left. \frac{\partial^2{e(\delta)}}{\partial {\delta}^2}
\right|_{\delta=0}.
\end{equation}
The GS order of the CCM model state will thus become unstable against
formation of the imposed form of order when $\chi \rightarrow \infty$
or, equivalently, when $1/\chi \rightarrow 0$.  The corresponding
LSUB$m$ results for the susceptibility of the given CCM model state
against the imposed form of order are then extrapolated to the
LSUB$\infty$ limit using the unbiased ``leading power-law'' scheme,
\begin{equation}
\chi^{-1}(m) = x_{0}+x_{1}(1/m)^{\nu}\,,            \label{Extrapo_inv-chi}
\end{equation}
similar to that in Eq.\ (\ref{M_extrapo_nu}) for the order parameter.

\begin{figure}[t]
\mbox{
\includegraphics[width=6cm,height=9cm,angle=270]{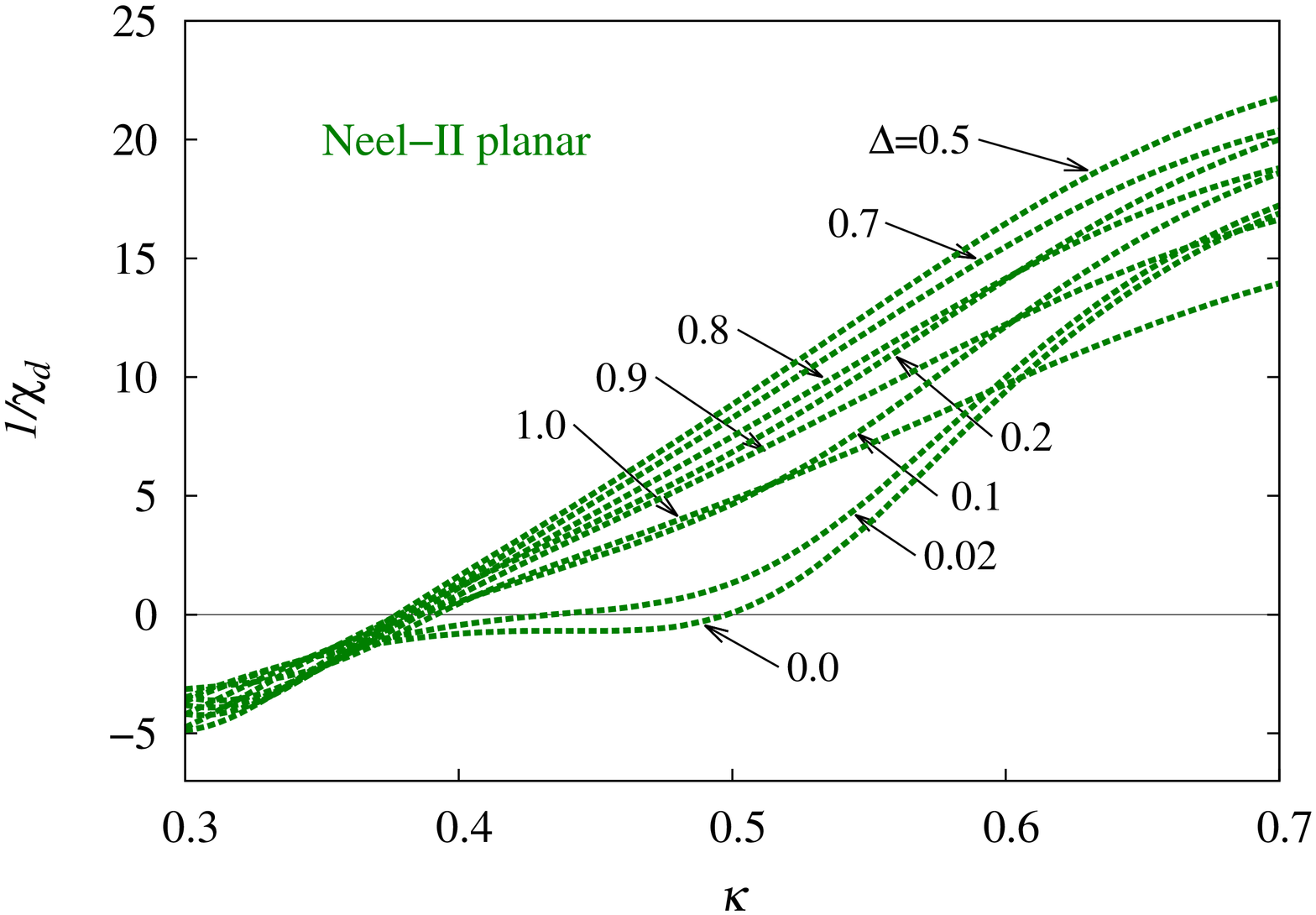}
\quad
\raisebox{-3.5cm}{
\includegraphics[width=2.2cm,height=2.2cm]{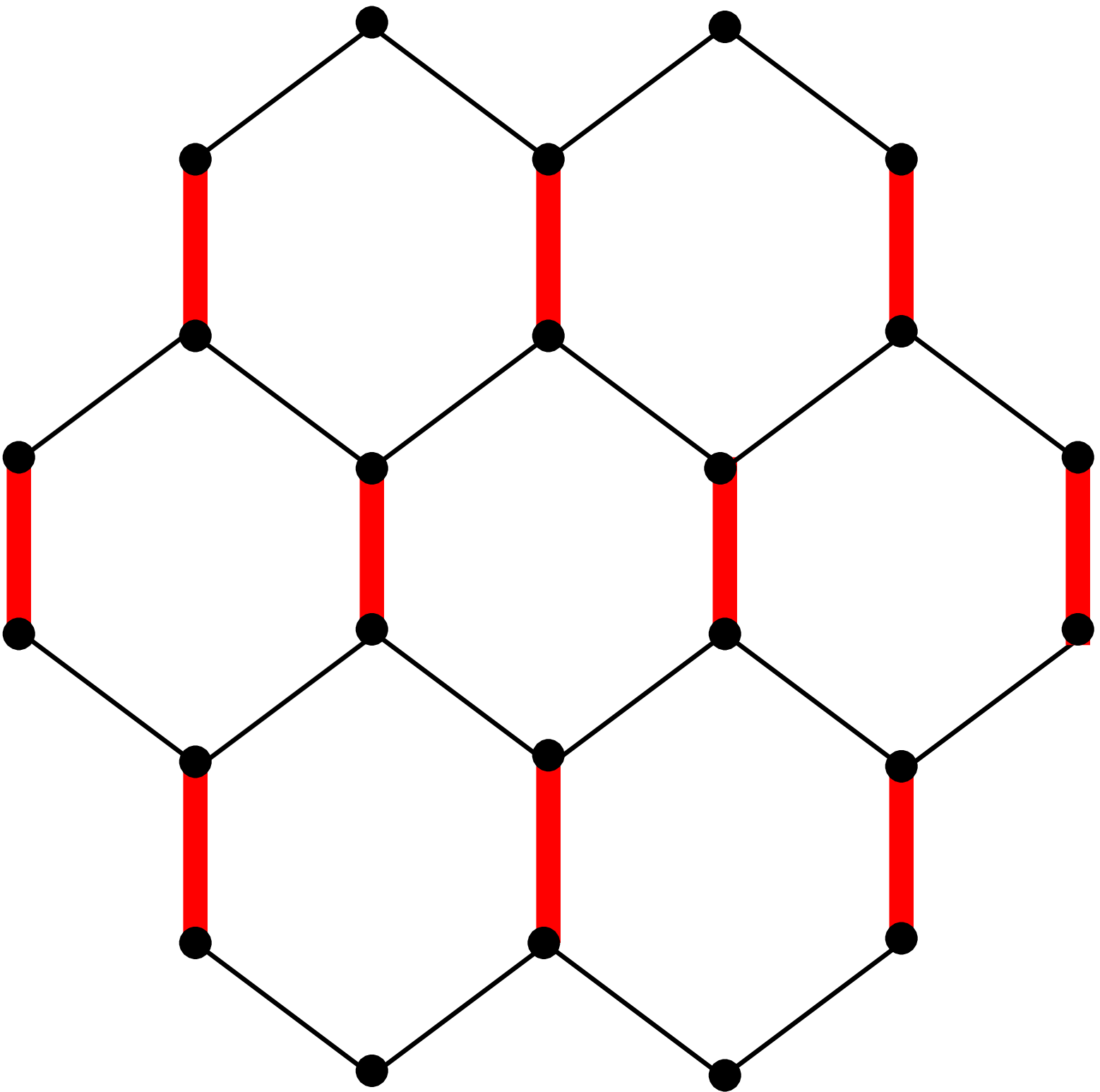}
}
}
\caption{(Color online) Left: The inverse staggered dimer susceptibility, $1/\chi_{d}$, versus the frustration parameter, $\kappa \equiv J_{2}/J_{1}$, for the
  spin-$\frac{1}{2}$ $J^{XXZ}_{1}$--$J^{XXZ}_{2}$ model on the honeycomb lattice (with $J_{1}=1$) for various values of the anisotropy parameter $\Delta$.  We show extrapolated CCM LSUB$\infty$ results (see text for details) based on the N\'{e}el-II planar state as CCM model state.  Right: The field $F \rightarrow
  \delta\; \hat{O}_{d}$ for the staggered dimer susceptibility,
  $\chi_{d}$.  Thick (red) and thin (black) lines correspond
  respectively to strengthened and unaltered NN exchange couplings,
  where $\hat{O}_{d} = \sum_{\langle i,j \rangle} a_{ij}(s^{x}_{i}{s}^{x}_{j}+s^{y}_{i}{s}^{y}_{j}+ \Delta s^{z}_{i}{s}^{z}_{j})$, and the sum runs over all NN
  bonds, with $a_{ij}=+1$ and 0 for thick (red) lines and thin
  (black) lines respectively.}
\label{X_SDVBC}
\end{figure}

Previous results using the CCM for the current model of Eq.\
(\ref{H_honey_XXZ}) in the limiting cases of the $XX$ model
\cite{Bishop:2014_honey_XY} at $\Delta=0$ and the $XXX$ model
\cite{Bishop:2012_honeyJ1-J2,RFB:2013_hcomb_SDVBC} at $\Delta=1$, as
well as those using alternative techniques, suggest that
N-II(p) ordering strongly competes with SDVBC ordering to form the
stable GS phase in the relevant part of phase space.  Hence, we now
perform CCM calculations based on the N-II(p) state as model state
where the perturbing field promotes SDVBC order, $\hat{O}
\rightarrow \hat{O}_{d}$, as illustrated schematically in the
right-hand frame of Fig.\ \ref{X_SDVBC}.
The results presented in Fig.\ \ref{X_SDVBC} for the corresponding
inverse staggered dimer susceptibility, $1/\chi_{d}$, are LSUB$\infty$
extrapolations based on Eq.\ (\ref{Extrapo_inv-chi}), with LSUB$m$
results $m=\{4,6,8\}$ used as input, for each of the values of
$\Delta$ shown.  They show clearly that the lower critical value of
the frustration parameter $\kappa$ at which SDVBC order appears is
rather insensitive to the value of the spin anisotropy parameter
$\Delta$ for all $\Delta \gtrsim 0.1$, where it takes the almost
constant value $\kappa \approx 0.38$.  However, the locus of such
SDVBC critical points meets the corresponding locus of critical points
$\kappa_{c_{u}}(\Delta)$ above which N-II(p) order appears, as taken
from Fig.\ \ref{M}, at a value $\Delta \approx 0.1$.  Hence, for
values $\Delta \lesssim 0.1$, a ``mixed'' region opens up in the $T=0$
GS phase diagram in which both SDVBC and N-II(p) forms of order appear
to coexist over a fairly narrow range of values of $\kappa$, above
which N-II(p) order then reasserts itself as the sole form of ordering
in the GS phase.

\begin{figure}[!b]
\mbox{
\includegraphics[angle=270,width=9cm]{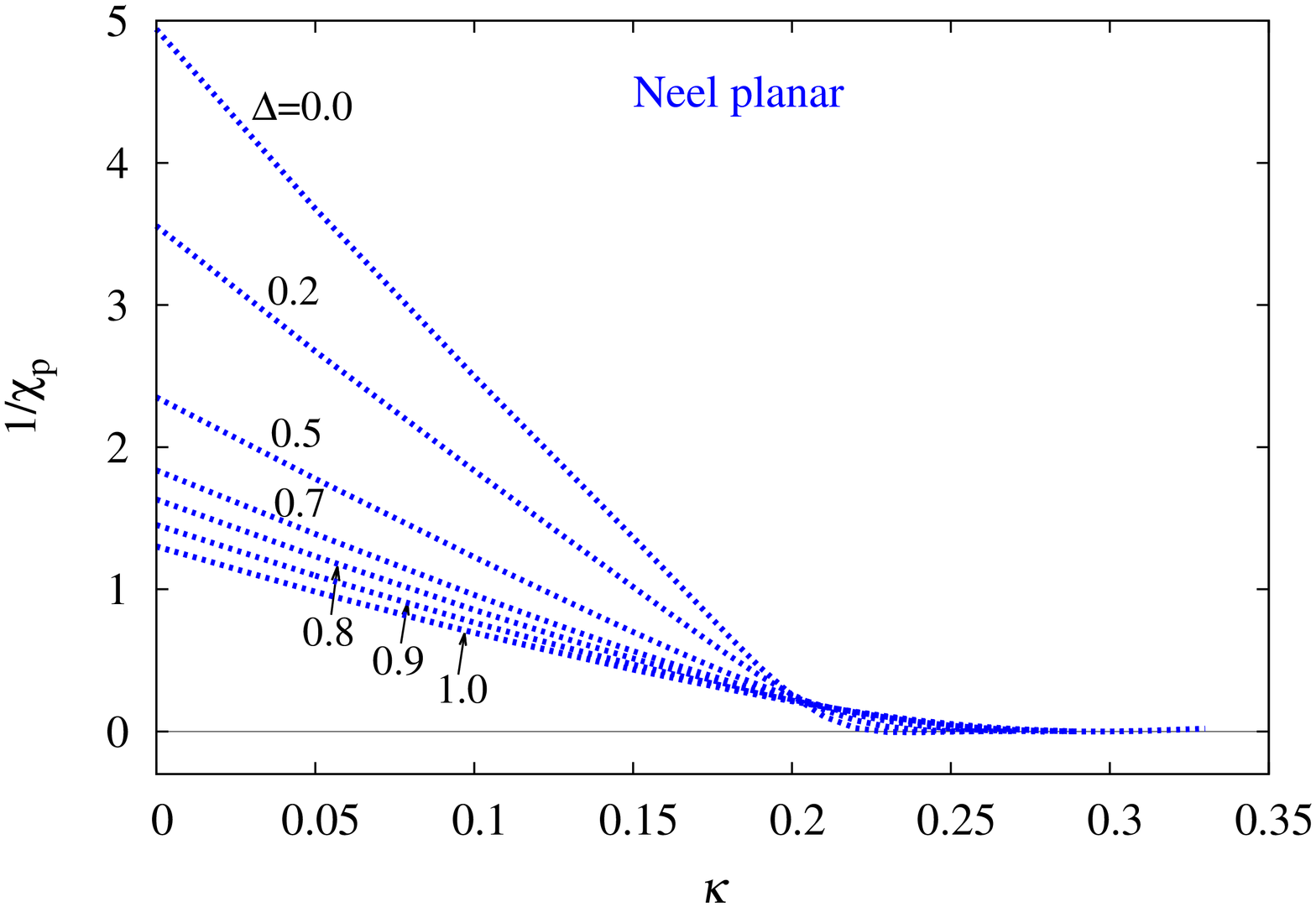}
\quad
\raisebox{-3.5cm}{
\includegraphics[width=2.2cm,height=2.2cm]{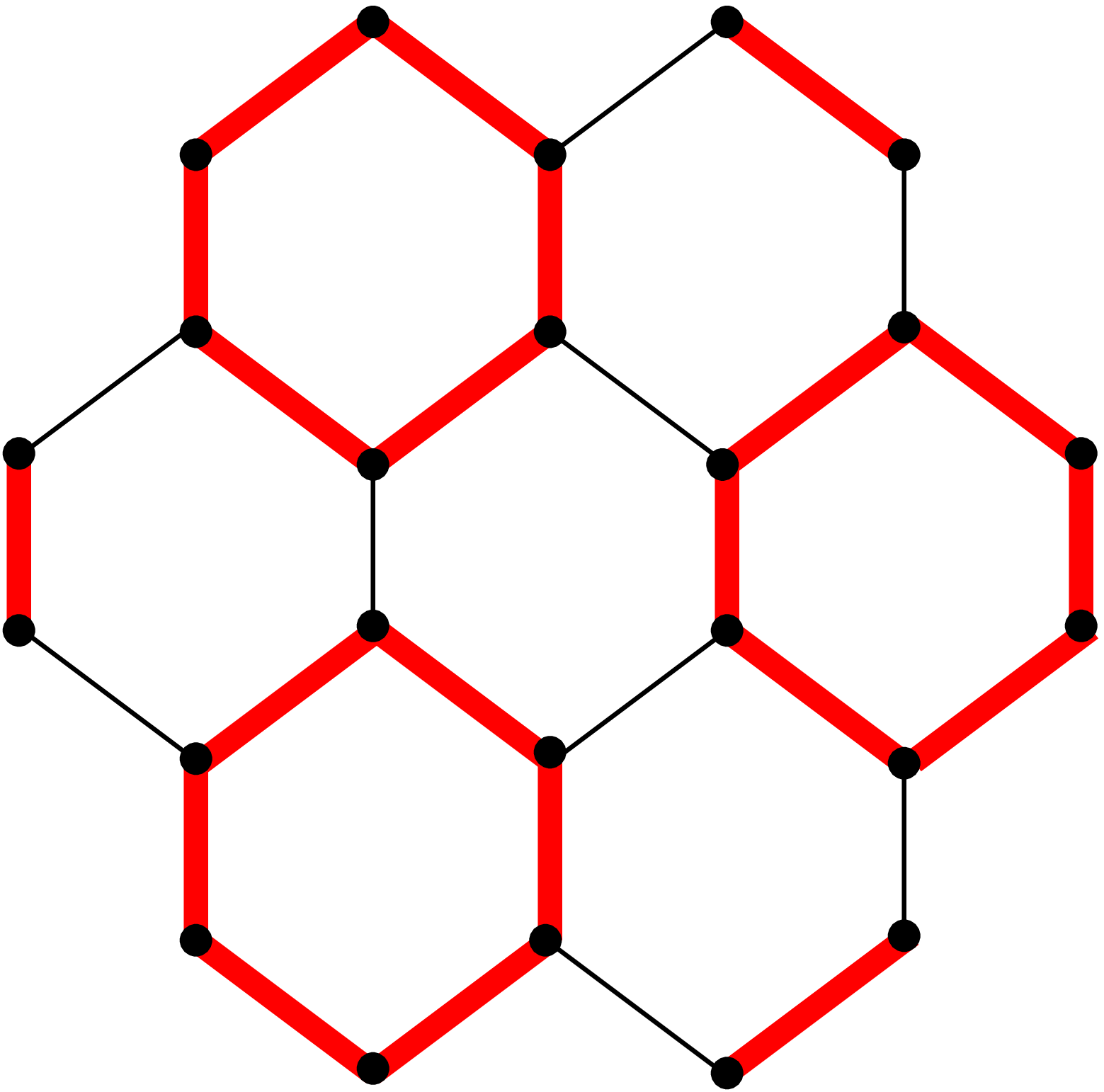}
}
}
\caption{(Color online) Left: The inverse plaquette susceptibility, $1/\chi_{p}$, versus the frustration parameter, $\kappa \equiv J_{2}/J_{1}$, for the
  spin-$\frac{1}{2}$ $J^{XXZ}_{1}$--$J^{XXZ}_{2}$ model on the honeycomb lattice (with $J_{1}=1$) for various values of the anisotropy parameter $\Delta$.  We show extrapolated CCM LSUB$\infty$ results (see text for details) based on the N\'{e}el planar state as CCM model state.  Right: The field $F=\delta\, \hat{O}_p$ for 
  the plaquette susceptibility $\chi_p$.  
Thick (red) 
  and thin (black) lines correspond respectively
  to strengthened and weakened NN exchange couplings, where $\hat{O}_{p} = \sum_{\langle i,j \rangle} a_{ij}(s^{x}_{i}{s}^{x}_{j}+s^{y}_{i}{s}^{y}_{j}+ \Delta s^{z}_{i}{s}^{z}_{j})$, and the sum runs over all NN
  bonds, with $a_{ij}=+1$ and $-1$ for thick (red) and thin (black)
  lines respectively.}
\label{PVBC}
\end{figure}

We turn finally to the remaining, and especially interesting, region
in the $\kappa$--$\Delta$ phase space, which is outside the region of
N(z) stability but between the two curves $\kappa =
\kappa_{c_{l}}(\Delta) \approx 0.21$ (below which N(p) order is
stable) and $\kappa \approx 0.38$ (above which SDVBC and/or N-II(p)
order is stable).  For the limiting case of the $XXX$ model (at
$\Delta=1$) some methods (including the CCM) favor the GS phase to
have PVBC order over all or part of this region
\cite{Albuquerque:2011_honey,
  Mosadeq:2011_honey,Li:2012_honey_full,Bishop:2012_honeyJ1-J2,RFB:2013_hcomb_SDVBC,Ganesh:2013_honey_J1J2mod-XXX,Zhu:2013_honey_J1J2mod-XXZ,Gong:2013_J1J2mod-XXX},
while others favor a QSL state
\cite{Clark:2011_honey,Mezzacapo:2012_honey,Gong:2013_J1J2mod-XXX,Yu:2014_honey_J1J2mod-XXZ},
again over all or part of the region.  Hence, we now perform CCM
calculations based on the N(p) state as model state, in the presence
of a perturbing field that now promotes PVBC order, $\hat{O}
\rightarrow \hat{O}_{p}$, as shown schematically in the right-hand
frame of Fig.\ \ref{PVBC}.
The results presented in Fig.\ \ref{PVBC} for the corresponding
inverse plaquette susceptibility, $1/\chi_{p}$, are again LSUB$\infty$
extrapolations based on Eq.\ (\ref{Extrapo_inv-chi}), with LSUB$m$
results $m=\{4,6,8\}$ used as input, for each of the values of
$\Delta$ shown.  Once again, they show clear evidence for
corresponding regions of stability of a GS phase with PVBC order.

\section{$T=0$ GS PHASE DIAGRAM}
\label{phase_sec}
\begin{figure}[!t]
  \includegraphics[angle=270,width=9cm]{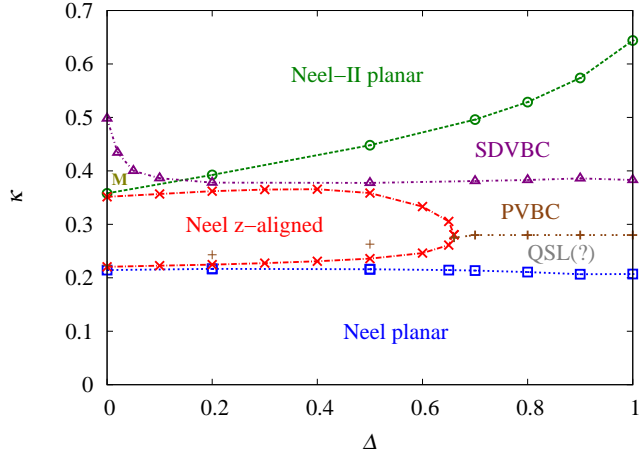}
  \caption{(Color online) Phase diagram for the spin-$\frac{1}{2}$
    $J^{XXZ}_{1}$--$J^{XXZ}_{2}$ model on the honeycomb lattice (with
    $J_{1}>0$ and $\kappa \equiv J_{2}/J_{1}>0$) in the window $0 \leq \kappa \leq 1$ and $0 \leq \Delta \leq 1$, as obtained by a CCM analysis.  The phase in the region marked ``M'' has both SDVBC and N\'{e}el-II planar order.  See text for details.}
\label{phase}
\end{figure}

On the basis of the results presented so far in Sec.\
\ref{results_sec} it is now straightforward to construct the $T=0$ GS
phase digram for the model, as shown in Fig.\ \ref{phase}.
Clearly, the regions of stability of the N(p), N($z$), and N-II(p)
phases may be taken from Fig.\ \ref{M} as those in which the
respective magnetic order parameters $M$ take positive values.  The
corresponding points at which $M=0$ are shown in Fig.\ \ref{phase} by
open square ($\square$), times ($\times$), and open circle
($\bigcirc$) symbols, respectively.  Similarly, the points at which
$\chi^{-1}_{d} \rightarrow 0$ and $\chi^{-1}_{p} \rightarrow 0$, taken
from Figs.\ \ref{X_SDVBC} and \ref{PVBC}, are shown in Fig.\
\ref{phase} by open triangle ($\triangle$) and plus ($+$) symbols,
respectively.  The small region of mixed SDVBC and N-II(p) order,
described in Sec.\ \ref{results_sec}, is denoted in Fig.\ \ref{phase} by ``M''.

Based on the results for $1/\chi_{p}$ from Fig.\ \ref{PVBC} we now
tentatively identify the region denoted by ``PVBC'' in Fig.\
\ref{phase} as having stable PVBC order.  The remaining region denoted
by ``QSL(?)'' is a clear candidate for a QSL phase, since we find no
evidence for any form of magnetic (spin) ordering, nor of either form
of VBC ordering, for which we have tested.  In this context we also
mention that a recent DMRG study \cite{Gong:2013_J1J2mod-XXX} of the
limiting $XXX$ case (i.e., $\Delta=1$) of the present model found
solid evidence of (weak) PVBC order in the thermodynamic ($N
\rightarrow \infty$) limit, in the range $0.26 \lesssim \kappa
\lesssim 0.35$ of the frustration parameter, in good agreement with our
own estimate for this limiting $XXX$ case that PVBC order exists in
the range $0.28 \lesssim \kappa \lesssim 0.38$.  Very interestingly,
the same DMRG study \cite{Gong:2013_J1J2mod-XXX} excluded, in the same
thermodynamic limit, any form of either magnetic (spin) or VBC
ordering in the range $0.22 \lesssim \kappa \lesssim 0.26$ immediately
above the N\'{e}el-ordered regime for the $XXX$ model, which was
identified as being the stable GS phase for $\kappa \lesssim 0.22$.
These DMRG findings were thus consistent with a QSL phase in the
region $0.22 \lesssim \kappa \lesssim 0.26$, again in broad agreement
with our own tentative conclusion of a QSL phase in the region $0.21
\lesssim \kappa \lesssim 0.28$ for the $XXX$ limiting case of the
model.  Indeed, these results are backed up by our earlier CCM
analysis \cite{Bishop:2012_honeyJ1-J2} of the $s=\frac{1}{2}$
$J_{1}$--$J_{2}$ $XXX$ model on the honeycomb lattice.

Thus, it was noted already in Ref.\ \cite{Bishop:2012_honeyJ1-J2} that
the transition from the N(p) phase to the PVBC phase in the $XXX$
model might be via an intermediate phase.  Any such intermediate phase
was estimated to be restricted to a region $\kappa_{c_{1}} < \kappa <
\kappa_{c_{1}}'$.  The value of $\kappa_{c_{1}}$ was accurately
obtained from the point where N\'{e}el order vanishes as
$\kappa_{c_{1}} = 0.207(3)$, and it is identical to that now shown in
Fig.\ \ref{phase} by the open square ($\square$) symbol at
$\Delta=1$.  The high accuracy obtained for $\kappa_{c_{1}}$
essentially stems from the shape of the N(p) order curve shown in
Fig.\ \ref{M}, with its very steep (or infinite) slope at the point
$\kappa_{c_{1}}$ where $M \rightarrow 0$.  By contrast, the point
$\kappa_{c_{1}}'$ was determined as in Fig.\ \ref{PVBC} from the point
where $1/\chi_{p} \rightarrow 0$.  The relative inaccuracy in this
value stems, conversely from the very shallow (or zero) slope in the
$1/\chi_{p}$ curve at the point $\kappa_{c_{1}}'$ where it becomes
zero.  In the earlier CCM analysis \cite{Bishop:2012_honeyJ1-J2} a
value $\kappa_{c_{1}}' \approx 0.24$ was quoted, without an error
estimation.  In the current analysis we have specifically examined the
lower phase boundary of the PVBC phase in greater detail, and our best
estimate for the limiting $XXX$ model is now $\kappa_{c_{1}}' \approx
0.28(2)$ from Fig.\ \ref{PVBC}, and as shown in Fig.\ \ref{phase} by
the plus ($+$) symbol at $\Delta=1$.  Nevertheless, it is still the case
that of all the phase boundaries shown in Fig.\ \ref{phase}, the one
between the PVBC and putative QSL phases probably has the largest
uncertainty, with a similar error along its whole length to that
quoted above at the point $\Delta=1$.  In this context we note too
that Fig.\ \ref{phase} shows that the plus ($+$) symbols denoting the
lower boundary of PVBC stability do not fall precisely on top of the
times ($\times$) symbols that denote the lower boundary of stability
of the N($z$) phase, in the region $\Delta \lesssim 0.66$ where the
latter phase exists as a stable GS phase.  This difference is probably
also another independent indication of the error bars associated with
the lower PVBC boundary points.

These error bars could certainly be reduced by including higher-order
LSUB$m$ results in the extrapolations.  The entire PVBC and SDVBC
regions of stability would also more definitively be confirmed by
performing calculations of $1/\chi_{p}$ and $1/\chi_{d}$ based on
other CCM model states to confirm their respective boundaries.  For
example, for the PVBC phase one might also use the N-II(p) state as
CCM model state to confirm the upper boundary of the phase.  In any
case, more definitive evidence awaits higher-order LSUB$m$
calculations.  Without them, for example, the possibility of a stable
QSL phase also existing in the very narrow region between the N($z$)
and SDVBC phases for $\Delta \lesssim 0.66$ also cannot be ruled out.

\section{CONCLUSIONS}
\label{conclusion}
In this paper we have outlined how the well-known CCM technique, which
has been very widely and very successfully applied to diverse (both
finite and macroscopically extended) physical systems that exist in a
spatial continuum, can be adapted for use with spin-lattice models of
interest in quantum magnetism, in which the spins are confined to the
sites of a regular periodic spatial lattice.  In particular, we have
explained how it may be applied, with comparable success, to high
orders in a systematically improvable hierarchy of approximations.
The method acts at every level of truncation in the thermodynamic
limit ($N \rightarrow \infty$), and the {\it only} approximation made
in practice is to a given $m$th level in the approximation hierarchy.
Thus, unlike in such alternative techniques as ED and QMC methods, no
finite-size scaling is ever needed within the CCM.  We have also shown
how GS quantities may readily be extrapolated to the exact $m
\rightarrow \infty$ limit of the truncation scheme, by the use of
well-tested heuristic schemes.

As an illustration of the CCM technique we applied it here to the
two-dimensional, frustrated, spin-half $J^{XXZ}_{1}$--$J^{XXZ}_{2}$
model on the honeycomb lattice.  We demonstrated explicitly how a CCM
analysis of the model could yield a fully coherent and accurate picture
of its full $T=0$ GS phase diagram.  We identified, in particular, a
specific region in the phase space in which we positively excluded
magnetic and VBC forms of order, and which is hence a
strong candidate for a QSL phase.  Clearly, it would be of value to
apply other techniques to this model in order to check our findings.

We note finally that the CCM has been applied with comparable success
in recent years to many other spin-lattice problems.  Particular
strengths of the method are that at every level of approximation it
obeys both the Goldstone linked-cluster theorem (in the sense that it
is manifestly size-extensive) and, perhaps even more importantly, the
Hellmann-Feynman theorem.

In conclusion, we hope that we have convinced the reader that the CCM is extremely versatile, requiring only the choice of a suitable model state (or set of such states) as input, on top of which the method incorporates the multispin correlations systematically.  Although we have demonstrated its use here for the case of a spin-half system, it is quite straightforward to generalize the CCM for use with spins of arbitrary quantum number $s$ \cite{Farnell:2002_1D}.


\begin{theacknowledgments}
We thank the University of Minnesota Supercomputing Institute for the
grant of supercomputing facilities.
\end{theacknowledgments}



\bibliographystyle{aipproc}   

\bibliography{bib_general}

\IfFileExists{\jobname.bbl}{}
 {\typeout{}
  \typeout{******************************************}
  \typeout{** Please run "bibtex \jobname" to optain}
  \typeout{** the bibliography and then re-run LaTeX}
  \typeout{** twice to fix the references!}
  \typeout{******************************************}
  \typeout{}
 }

\end{document}